\documentclass[a4paper,english,11pt,amssymb,nofootinbib,superscriptaddress]{revtex4-2}
\usepackage{lmodern}
\usepackage{lmodern}

\usepackage[T1]{fontenc}
\usepackage[latin9]{inputenc}
\usepackage{babel}
\usepackage{mathrsfs}
\usepackage{amsmath}
\usepackage{amssymb}
\usepackage{esint}
\usepackage{dsfont}
\usepackage[unicode=true,pdfusetitle,
bookmarks=true,bookmarksnumbered=false,bookmarksopen=false,
breaklinks=false,pdfborder={0 0 1},backref=false,colorlinks=false]
{hyperref}
\usepackage{xcolor}

\makeatletter

\@ifundefined{textcolor}{}
{%
	\definecolor{BLACK}{gray}{0}
	\definecolor{WHITE}{gray}{1}
	\definecolor{RED}{rgb}{1,0,0}
	\definecolor{GREEN}{rgb}{0,1,0}
	\definecolor{BLUE}{rgb}{0,0,1}
	\definecolor{CYAN}{cmyk}{1,0,0,0}
	\definecolor{MAGENTA}{cmyk}{0,1,0,0}
	\definecolor{YELLOW}{cmyk}{0,0,1,0}
}

\usepackage{babel}
\usepackage{babel}
\usepackage{babel}
\usepackage{babel}
\usepackage{babel}
\usepackage{babel}
\usepackage{babel}
\usepackage{babel}
\usepackage{babel}
\usepackage{babel}
\usepackage{babel}
\usepackage{babel}
\usepackage{babel}
\usepackage{babel}
\usepackage{babel}
\usepackage{babel}
\usepackage{babel}
\usepackage{graphicx}
\def\b{\begin{equation}}
\def\e{\end{equation}}

\@ifundefined{textcolor}{}{%
	\definecolor{BLACK}{gray}{0}
	\definecolor{WHITE}{gray}{1}
	\definecolor{RED}{rgb}{1,0,0}
	\definecolor{GREEN}{rgb}{0,1,0}
	\definecolor{BLUE}{rgb}{0,0,1}
	\definecolor{CYAN}{cmyk}{1,0,0,0}
	\definecolor{MAGENTA}{cmyk}{0,1,0,0}
	\definecolor{YELLOW}{cmyk}{0,0,1,0}
}

\usepackage{latexsym}\usepackage{bm}

\makeatother

\begin{document}
	\title{Refined thermodynamics of black holes with proper conserved charges}
	
	\author{Aydin Tavlayan}
	
	\email{aydint@metu.edu.tr}
	
	\selectlanguage{english}%
	
	\affiliation{Department of Physics,\\
		Middle East Technical University, 06800 Ankara, Turkey}
	\author{Bayram Tekin}
	
	\email{btekin@metu.edu.tr}
	\affiliation{Department of Physics,\\
		Middle East Technical University, 06800 Ankara, Turkey}

	\selectlanguage{english}%
\begin{abstract}
Field equations of a classical, geometric, theory of gravity, augmented with some semiclassical considerations strongly suggest that the gravitational field representing a stationary black hole can be simply described with a few thermodynamical coordinates and their conjugates that obey the four laws of thermodynamics plus the Smarr formula and the reverse isoperimetric inequality that bounds the maximum entropy for a given effective volume of space. The thermodynamics of black holes is a promising window to the quantum nature of black holes; hence, it is important to understand all the details of these laws. The identification and the meaning of these thermodynamic coordinates depend on the gravity theory under consideration. For example, the existence of dimensionful coupling constants, such as the cosmological constant, changes the scaling properties of the theory, its solutions, and the laws of thermodynamics. Here we show, using the background Killing charge method, which applies to the black hole solutions of any gravity theory that has a maximally symmetric vacuum, to define the mass and angular momentum, instead of using the Komar mass and the angular momentum, how the thermodynamics of black holes such as the $D$ dimensional Kerr-AdS black holes in cosmological Einstein's theory and the spherically symmetric black holes in the Einstein-Gauss-Bonnet theory changes. We give the effective volume of black holes even without a cosmological constant.
\end{abstract}
\maketitle
\section{Introduction} 
Quantum mechanics was born as a continuation of thermodynamics; an arduous process best exemplified by the long search of understanding the details, such as the exact spectrum of the black-body radiation. Fully understanding the fundamental physics of the microscopic degrees of freedom that emit and absorb electromagnetic waves from a heated object required almost a century of work between the 1860s to 1960s, when quantum electrodynamics was established as a consistent theory of light and charged particles. Since the 1970s we are, in some sense, in a similar, albeit a much harder path: we are trying to understand the thermodynamics of a black-body that also self-gravitates, the case which necessarily forces us to understand the thermodynamics of the gravitational field.  If the history rhymes, one hopes that after formulating consistent macroscopic thermodynamics for the gravitational field, one will also understand the statistical origins of these laws and eventually arrive at the quantum nature of gravity which presumably would involve identifying the microscopic degrees of freedom responsible for generating gravity. At the moment, we are far from this goal. Yet, so far, much has been accomplished in this vein: for example, we now understand that a quantum mechanical temperature that is proportional to $\hbar$ times the surface gravity can be assigned to a quasistationary black hole \cite{Hawking1}; and one can associate an entropy with an arbitrary portion of a gravitational field both in Einstein's gravity \cite{Visser1} and in higher curvature theories \cite{Tavlayan1}, in the same way that the entropy was assigned to black holes \cite{Bekenstein1} or de Sitter spacetimes \cite{Gibbons1}. In Einstein's gravity, besides the four laws of black hole mechanics, three of which are relations between the infinitesimal thermodynamical coordinates and their conjugates, one also has the Smarr relation \cite{Smarr1} that arises as a result of dimensional analysis and connects the finite versions of these coordinates, which is akin to the Gibbs-Duhem relation in classical thermodynamics. Even though we only have theoretical arguments, the fact that some gravitational fields, such as those describing stationary black holes, obey the laws of thermodynamics is rather remarkable and by no means obvious because some of these solutions are pure geometry with no matter fields.

Einstein's gravity is expected to receive higher curvature corrections at small scales,  most relevant to singularities of black holes and cosmological singularities. Understanding the refined thermodynamics of the gravitational field in these better-UV-behaved theories is crucial for quantum gravity. For example, the Smarr relation often fails when one considers these extended theories of gravity just like the laws of black hole thermodynamics unless one accommodates the corresponding corrections associated with the coupling constants that appear in each added term in the action in a proper way \cite{Hajian1, Kastor2}.  The crux of the matter here is the following: thermodynamical coordinates approximate the gravitational field in some sense, and must be defined properly not just as geometric objects computed from the metric or the geometry alone, but computed in a given theory.  For example, the mass and angular momenta can be defined as Komar charges in a purely geometric way \cite{Komar1} or, as conserved charges in a given theory \cite{Abbott1, Deser1, Deser2} corresponding to asymptotic or background symmetries. Here we follow the latter method: given a generic gravity theory defined by the covariant field equations $\cal{E}_{\mu \nu} = \kappa T_{\mu \nu}$, one first searches for a vacuum solution (i.e., the background)  $(\bar{g}_{\mu \nu})$  which is maximally symmetric with Killing vectors $\bar \xi^\mu$. Then for any other solution $g_{\mu \nu}$, be it a vacuum or nonvacuum solution, that asymptotically locally approaches the background solution, one can construct the {\it partially } conserved current $j^\mu := \sqrt{|\bar g|}{\cal{E}}_L^{\mu \nu} \bar \xi_\nu$. Integration of the zeroth component over the spatial volume yields the corresponding Killing charge $Q[\bar \xi] := \int_{\bar\Sigma}  d^{(D-1)}x  \sqrt{|\bar g|}{\cal{E}}_L^{0  \nu} \bar \xi_\nu$. Furthermore, this integral can be reduced to a surface integral over $\partial \bar \Sigma$: For cosmological Einstein gravity, one arrives at the formula (\ref{newcharge}).  The versatility of the model is clear: unlike the Komar charge definition that has a different numerical factor (i.e., Komar anomaly) for the mass and angular momentum, there is a single formula for all conserved charges; and the charge expression can easily be found for any geometric theory of gravity.  One important caveat is the following: the background spacetime is assumed to have zero conserved charges in this method.  Using these expressions we establish the following: First, we show that the laws of black hole thermodynamics and the Smarr relation are satisfied for both the cosmological Einstein \cite{Hawking2, Pope1, Pope2, Myers1} and the Einstein-Gauss-Bonnet (EGB) theory. Then, we show that these quantities respect the geometry such as the reverse isoperimetric inequality \cite{Cvetic1}. Finally, we concentrate on the EGB theory which has a dimensionful coupling constant and show that, by adding proper terms associated with it, new definitions of mass and angular momenta satisfy thermodynamics laws and the Smarr relation. The procedure outlined here can be extended to any gravity theory that admits a maximally symmetric spacetime as a solution as long as the black hole solutions are known. 

\section{The Thermodynamics of Einstein Gravity with a cosmological constant}
As the generic $D \ge 4$ dimensional computation becomes somewhat cumbersome, we first establish the definitions in the five-dimensional (5D) rotating black hole with two rotation parameters and then give the results of the generic dimensional cases.

\subsection{Kerr-AdS black hole in five dimensions}
In the units
$\hbar = c = k_B = G = 1$ and $\Omega_{3} = 2 \pi^2$, the theory defined by the action
\begin{equation}
I=\frac{1}{4 \Omega_{3}} \int d^5x \sqrt{-g} \left(R-2\Lambda\right)
\end{equation}
admits a three-parameter solution (the rotation parameters $a$, $b$ and the mass parameter $m$), which in the coordinates $x^{\mu}=\left(t,r,\theta,\varphi,\psi\right)$ reads  \cite{Hawking2,Pope1,Pope2} 
\begin{eqnarray}\label{5Dmetric}
ds^2&=& - \frac{\Delta_{\theta} \left(1-\frac{\Lambda r^2}{6}\right)}{\Xi_a \Xi_b}dt^2 + \frac{\rho^2}{\Delta_{r}}dr^2 + \frac{\rho^2}{\Delta_{\theta}}d\theta^2\\ 
&&+\frac{r^2+a^2}{\Xi_a}\sin^2{\theta} d\varphi^2 + \frac{r^2 + b^2}{\Xi_b}\cos^2{\theta} d\psi^2 \nonumber\\
&&+\frac{2m}{\rho^2}\left(\frac{\Delta_{\theta}}{\Xi_a \Xi_b}dt - \frac{a^2 \sin^2{\theta}}{\Xi_a}d\varphi - \frac{b^2 \cos^2{\theta}}{\Xi_b}d\psi\right)^2,\nonumber
\end{eqnarray}
where the metric functions are 
\begin{eqnarray}
&&\Delta_r := \frac{\left(r^2+a^2\right)\left(r^2+b^2\right)\left(1-\frac{\Lambda r^2}{6}\right)}{r^2}-2m, \nonumber\\
&&\Delta_{\theta}:=1+\frac{a^2 \Lambda}{6}\cos^2{\theta} + \frac{b^2 \Lambda}{6}\sin^2{\theta},\nonumber\\
&&\rho^2 := r^2 + a^2 \cos^2{\theta} + b^2 \sin^2{\theta},
\end{eqnarray}
and $\Xi_a = 1+ \frac{a^2\Lambda}{6}$, $\Xi_b = 1+ \frac{b^2\Lambda}{6}$. The (outer) event horizon is located at $r_{\mathrm{H}}$ at which the Killing vector $\xi_ {\mathrm{H}}= \partial_t + \Omega_{\mathrm{H}}^\varphi\partial_\varphi+ \Omega_{\mathrm{H}}^\psi\partial_\psi$ becomes null with the horizon angular velocities given as 
\begin{equation}
\Omega_{\mathrm{H}}^{\varphi}  =\frac{a\left(1-\frac{\Lambda r_{\mathrm{H}}^2}{6}\right)}{\left(r_{\mathrm{H}}^2+a^2\right)}, \quad \Omega_{\mathrm{H}}^\psi=\frac{b\left(1-\frac{\Lambda r_{\mathrm{H}}^2}{6}\right)}{\left(r_{\mathrm{H}}^2+b^2\right)}.\label{coord1}
\end{equation}
So $r_{\mathrm{H}}$ is given by the largest root of 
\begin{equation}
(1-\frac{\Lambda}{6}r_{\mathrm{H}}^2)(r_{\mathrm{H}}^2+a^2)(r_{\mathrm{H}}^2+b^2)-2m {r_\mathrm{H}}^2=0.  
\end{equation}
These are sufficient to calculate the thermodynamical quantities of the solution. The entropy of this 5D rotating black hole is
\begin{equation}
S=\frac{\pi\left(r_{\mathrm{H}}^2+a^2\right)\left(r_{\mathrm{H}}^2+b^2\right)}{\Xi_a \Xi_b r_{\mathrm{H}}},\label{coord2}
\end{equation}
while its temperature is \cite{Hajian2}
\begin{equation}
T  =\frac{r_{\mathrm{H}}^4\left(1-\frac{\Lambda}{6}\left(2 r_{\mathrm{H}}^2+a^2+b^2\right)\right)-a^2 b^2}{2 \pi r_{\mathrm{H}}\left(r_{\mathrm{H}}^2+a^2\right)\left(r_{\mathrm{H}}^2+b^2\right)}.\label{coord3}
\end{equation}
Please note that, due to the overall normalization of our action, the area law reads as $S= \frac{\pi}{\Omega_{D-2}}A$.
 To properly account for the cosmological constant in black hole thermodynamics, following \cite{Henneaux1}, a rank-four totally antisymmetric gauge field is introduced \cite{Hajian2} in such a way that the cosmological constant appears as a parameter of the solution and not a coupling constant defining the theory. This is also well-motivated in string or supergravity theories in which the cosmological constant appears as an expectation value of some potential, or appears after compactification \cite{Cvetic1}.
 The desired action reads 
\begin{equation}
I=\int \frac{d^5 x}{16 \pi}\left(R \mp \frac{2}{5 !} F^2 \pm \frac{4 \nabla_\mu\left(A_{\mu_2 \ldots \mu_5} F^{\mu \mu_2 \ldots \mu_5}\right)}{4!}\right),
\end{equation}
which has a $U(1)$ gauge symmetry: $A \rightarrow A+ d\lambda(x)$ where $\lambda(x)$ is a totally antisymmetric three-form. The global part of this gauge symmetry yields, via the usual Noether's theorem, a conserved charge that is called the ''cosmological charge'' given as  \cite{Chernyavsky1}
\begin{equation}
C:=\pm\frac{\sqrt{\vert\Lambda\vert}}{2 \pi^2}.
\label{coord4}
\end{equation}
For this conserved charge, there exists a thermodynamical conjugate, the ''cosmological potential'' defined as a surface integral of the contraction of the hypersurface-generating null-Killing vector field with the gauge potential as
\begin{equation}
\Theta_{\mathrm{H}}:= \oint_{\mathrm{H}} \xi_{\mathrm{H}} \cdot A,
\end{equation}
where $A$ is the cosmological gauge field $A_{\mu_1}\ldots_{\mu_{4}}$. This integral, for the 5D metric above, yields the cosmological potential
\begin{align}
\Theta_{\mathrm{H}}=-\frac{\sqrt{|\Lambda|} \pi^2}{\Xi_a \Xi_b}&\left(\frac{\left(r_{\mathrm{H}}^2+a^2\right)\left(r_{\mathrm{H}}^2+b^2\right)}{2}\right.\nonumber\\
&\left.+\frac{m\left(a^2+b^2+\frac{\Lambda a^2 b^2}{3}\right)}{3 \Xi_a \Xi_b}\right). \label{coord5}
\end{align}
The mass and angular momenta of the solution can be computed from the following formula \cite{New1, New2} (which is given for generic $D$ spacetime dimensions as is needed later):
\begin{align}
Q\left(\bar{\xi}\right)=k\int_{\partial\bar{\Sigma}}d^{D-2}x\,\sqrt{\bar{\gamma}}\,\bar{\epsilon}_{\mu\nu}\left(R^{\nu\mu}\thinspace_{\beta\sigma}\right)^{\left(1\right)}\bar{\text{\ensuremath{{\cal {F}}}}}^{\beta\sigma},\label{newcharge}
\end{align}
where $k:= \frac{(D-1)(D-2)}{8(D-3)\Lambda\Omega_{D-2}}$ and
$(R^{\nu\mu}\thinspace_{\beta\sigma})^{\left(1\right)}$ is
the linearized part of the Riemann tensor about the (A)dS background;  while the antisymmetric two-form reads as $\bar{{\cal {F}}}^{\beta\sigma}=\bar{\nabla}^{\beta}\bar{\xi}^{\sigma}$ with 
$\bar{\xi}^{\sigma}$  being a background Killing vector corresponding to the relevant symmetry. The antisymmetric two-form $\bar \epsilon$ has the components $\bar{\epsilon}_{\mu\nu}:=\frac{1}{2}\left(\bar{n}_{\mu}\bar{\sigma}_{\nu}-\bar{n}_{\nu}\bar{\sigma}_{\mu}\right)$,
 where $\bar{n}_{\mu}$ is a normal one-form on the boundary of spacetime, $\partial\bar{\mathscr{M}}$,
 $\bar{\sigma}_{\nu}$ is the unit normal one-form on the spatial boundary $\partial\bar{\Sigma}$,
and $\bar{\gamma}$ is the induced metric on the boundary. For the metric (\ref{5Dmetric}), with $\bar{\xi}= - \partial_t$, $\bar{\xi}= \partial_\varphi$, and $\bar{\xi}= \partial_\psi $, one obtains, respectively \cite{Deser3}, the following conserved charges:
\begin{equation}\label{coord6}
E=\frac{ m\left(2 \Xi_a+2 \Xi_b-\Xi_a \Xi_b\right)}{2 \Xi_a^2 \Xi_b^2}, \, J_{\varphi}=\frac{ a m}{ \Xi_a^2 \Xi_b}, \, J_\psi=\frac{b m}{ \Xi_b^2 \Xi_a}.
\end{equation}
Finally, using these thermodynamical coordinates and their conjugates (\ref{coord1}), (\ref{coord2}), (\ref{coord3}), (\ref{coord4}), (\ref{coord5}), (\ref{coord6}), one arrives at the first law of black hole thermodynamics: 
\begin{eqnarray}
\delta E=T \delta S+\Omega_{\mathrm{H}}^{\varphi} \delta J_{\varphi}+\Omega_{\mathrm{H}}^\psi \delta J_\psi+\Theta_{\mathrm{H}} \delta C, 
\end{eqnarray}
and the following Smarr relation is satisfied:
\begin{eqnarray}
 E=3 T S+3 \Omega_{\mathrm{H}}^{\varphi} J_{\varphi}+3 \Omega_{\mathrm{H}}^\psi J_\psi-\Theta_{\mathrm{H}} C.
\end{eqnarray}
It is apt to note the following: with the necessary introduction of the cosmological constant as a conserved charge of the solution, the physical meaning $E$ as energy changes: now it is the {\it gravitational enthalpy} of the solution, i.e., the energy needed to create this black hole with the given properties in a spacetime with the given cosmological constant. The cosmological constant (or here the cosmological charge) and the pressure can be identified while the cosmological potential defines an effective volume of the black hole that can be interpreted as the volume inside the event horizon, or the difference of the renormalized volume of total spacetime when the difference refers to the difference of regularized volumes in the presence and the absence of the black hole \cite{Brown2, Caldarelli1}. Therefore, the missing $P \delta V$ term in the conventional thermodynamics arises as $V \delta P$ in the black hole physics. This interpretation and the related thermodynamical volume of black holes has given rise to an interesting subbranch of black hole physics dubbed ''black hole chemistry'' \cite{Caldarelli1, Kastor1, Kubiznak1, Kubiznak2, Dolan1}. Setting the negative heat capacity of black holes aside, the fact that these thermodynamical relations are satisfied by the pure gravitational field, that is, pure geometry, in a way similar to the thermodynamical relations of matter, gives one the following hint: thermodynamics, as our most crude description of a physical system, captures the gross features of a large cloud of gas not only before but also after it collapses into a black hole.  
Next, we generalize these results to generic $D\ge 4$ dimensions. The discussion bifurcates because of the special features of odd and even dimensions that allow different numbers of integration parameters. As the metric is somewhat cumbersome, we give the details of the computations in Appendix A and note the salient features here. 

\subsection{Kerr-AdS spacetimes in even dimensions}
Taking  the action to be $I=\frac{1}{4\Omega_{D-2}}\int d^Dx\sqrt{-g}\left(R-2\Lambda\right)$,
with an even number of dimensions, $D=2N+2$, there are $N$ distinct rotations; and the thermodynamic quantities become
\begin{eqnarray}
&&E=\frac{m}{\Xi} \sum_{i=1}^N \frac{1}{\Xi_i}, \quad J_i=\frac{m a_i}{\Xi_i \Xi},\nonumber\\
&&T=\frac{r_{\mathrm{H}}\left(1+g^2 r_{\mathrm{H}}^2\right)}{2 \pi} \sum_{i=1}^N \frac{1}{r_{\mathrm{H}}^2+a_i^2}-\frac{1-g^2 r_{\mathrm{H}}^2}{4 \pi r_{\mathrm{H}}},\nonumber\\
&&S=\pi\prod_{i=1}^N \frac{r_{\mathrm{H}}^2+a_{i}^2}{\Xi_i}, \quad \Omega_i=\frac{\left(1+g^2 r_{\mathrm{H}}^2\right) a_i}{r_{\mathrm{H}}^2+a_i^2},\nonumber\\
&&C = \pm \frac{\sqrt{\vert \Lambda \vert}}{\Omega_{D-2}},
\end{eqnarray}
where $\Xi_i := 1-g^2 a_i^2$, $g^2 := -\frac{2 \Lambda}{\left(D-1\right)\left(D-2\right)}$, and $\Xi := \prod_{i=1}^N \Xi_i$.
The event horizon is located at the largest root of the polynomial
\begin{equation}
\left(1+g^2 r_{\mathrm{H}}^2\right) \prod_{i=1}^N \left(r_{\mathrm{H}}^2+a_i^2\right)-2 mr_{\mathrm{H}}=0. 
\end{equation}
These quantities satisfy the first law of black hole thermodynamics 
\begin{equation}
\delta E=T \delta S+\sum_{i=1}^N \Omega_i \delta J_i+\Theta_{\mathrm{H}} \delta C \label{firstlaw}
\end{equation}
and the Smarr relation
\begin{equation}
\left(D-3\right) E = \left(D-2\right) \left(T S + \sum_{i=1}^N \Omega_i J_i \right)- \Theta_{\mathrm{H}} C. \label{Smarr}
\end{equation}

\subsection{The static limit} 
It pays to understand the no-rotation limit (the Schwarzschild-Tangherlini limit) and derive the isoperimetric ratio. Note that we still keep a nonzero cosmological constant.
In the static limit, i.e., $a_i \rightarrow 0$, the thermodynamic quantities become
\begin{eqnarray}
&&\Xi_i = 1, \quad \Xi = 1, \quad J_i=0, \quad \Omega_i=0, \nonumber\\
&& E= m \left(\frac{D-2}{2}\right), \quad S= \pi r_{\mathrm{H}}^{D-2}, \nonumber\\
&& T=\frac{r_{\mathrm{H}}\left(1+g^2 r_{\mathrm{H}}^2\right)}{2 \pi} \sum_{i=1}^N \frac{1}{r_{\mathrm{H}}^2}-\frac{1-g^2 r_{\mathrm{H}}^2}{4 \pi r_{\mathrm{H}}} \nonumber\\
&&=\frac{1}{4 \pi r_{\mathrm{H}}}\bigg(\left(1+g^2 r_{\mathrm{H}}^2\right)\left(D-2\right)-\left(1-g^2 r_{\mathrm{H}}^2\right)\bigg ).
\end{eqnarray}
From the Smarr law, one obtains
\begin{eqnarray}
&&\left(D-3\right) E = \left(D-2\right) T S - \Theta_{\mathrm{H}} C,\\
&&\frac{m}{2} \left(D-2\right) \left(D-3\right) = \frac{\left(D-2\right)r_{\mathrm{H}}^{D-3}}{4}\nonumber\\
&&\times\bigg(\left(1+g^2 r_{\mathrm{H}}^2\right)\left(D-2\right)-\left(1-g^2 r_{\mathrm{H}}^2\right)\bigg)-\Theta_{\mathrm{H}} C.\nonumber
\end{eqnarray}
By using the polynomial equation defining the event horizon, $m =\frac{r_{\mathrm{H}}^{D-3}}{2}\left(1+g^2 r_{\mathrm{H}}^2\right)$, one gets
\begin{eqnarray}
&&\Theta_{\mathrm{H}} C = - \frac{1}{4}r_{\mathrm{H}}^{D-3}\left(1+g^2 r_{\mathrm{H}}^2\right) \left(D-2\right) \left(D-3\right) \nonumber\\
&&+ \frac{\left(D-2\right) r_{\mathrm{H}}^{D-3}}{4}\left[\left(1+g^2 r_{\mathrm{H}}^2\right)\left(D-2\right)-\left(1-g^2 r_{\mathrm{H}}^2\right)\right],\nonumber\\
&&= - \frac{r_{\mathrm{H}}^{D-1} \Lambda }{\left(D-1\right)}.
\end{eqnarray}
From the definition of the cosmological charge $C=-\frac{\sqrt{\vert\Lambda\vert}}{\Omega_{D-2}}$, one can see that the cosmological potential reads as
\begin{equation}
\Theta_{\mathrm{H}} = -\frac{\sqrt{\vert\Lambda\vert} \Omega_{D-2} r_{\mathrm{H}}^{D-1} }{D-1}.
\end{equation}
Using the definition of the effective thermodynamic volume,  $\Theta_{\mathrm{H}} = - \sqrt{\vert\Lambda\vert} V_{\text{eff}}$, one gets an effective volume for the $D$-dimensional Schwarzschild-AdS black holes
\begin{equation}
V_{\text{eff}} = \frac{\Omega_{D-2}r_{\mathrm{H}}^{D-1}}{D-1}, \label{24}
\end{equation}
which does not make sense in the usual black hole thermodynamics without a cosmological constant. But, one can argue that  $V_{\text{eff}}$  survives in the vanishing cosmological constant limit, as the volume of the Schwarzschild black hole. 

\subsection{Reverse isoperimetric inequality}
For the static black hole, since the area of the event horizon is $A  = \Omega_{D-2} r_{\mathrm{H}}^{D-2}$, one can show that the reverse isoperimetric inequality is saturated 
\begin{equation}
\bigg(\frac{\left(D-1\right)V_{\text{eff}}}{\Omega_{D-2} }\bigg)^{\frac{1}{D-1}} \ge \bigg(\frac{A}{\Omega_{D-2}}\bigg)^{\frac{1}{D-2}},\label{25}
\end{equation}
as was expected for Schwarzschild-AdS black holes \cite{Cvetic1}. Interestingly, for a given volume $V_{\text{eff}}$, the bound is saturated for only the Schwarzschild-AdS black holes, and for any other solution, one obtains a reverse isoperimetric relation meaning that the Schwarzschild-AdS black hole has the maximum entropy among the same volume spaces. For example, the introduction of charge or rotation reduces the entropy-to-volume ratio \cite{Cvetic1}. In the rotating case for an even number of dimensions, one can show that the effective volume reads
\begin{equation}
    V_{\text{eff}}=\frac{m \Omega_{D-2}}{\Lambda \Xi}\left(-\sum_{i=1}^N \frac{1}{\Xi_i}+\frac{\left(D-2\right)}{2}\frac{\left(1-g^2 r_{\mathrm{H}}^2\right)}{1+g^2 r_{\mathrm{H}}^2}\right),\label{26}
\end{equation}
and the event horizon area reads as
\begin{equation}
    A=\Omega_{D-2}\prod_{i=1}^N \frac{r_{\mathrm{H}}^2+a_{i}^2}{\Xi_i}.\label{27}
\end{equation}
The effective volume, (\ref{26}), and the event horizon area, (\ref{27}), satisfy the inequality given in (\ref{25}). Observe that one can take the $\Lambda \rightarrow 0$ limit of  (\ref{26}) to define the volume of asymptotically flat black holes in even dimensions as 
\begin{equation}
    V_{\text{eff}}=\frac{2m \Omega_{D-2}r_{\mathrm{H}}^2}{D-1}, \label{28}
\end{equation}
where $r_{\mathrm{H}}$ is the largest root of $ \prod_{i=1}^N \left(r_{\mathrm{H}}^2+a_i^2\right)-2 mr_{\mathrm{H}}=0$ and the effective volume reduces to the Schwarzschild case (\ref{24}) when all $a_i$ vanish.

\subsection{Kerr-Ads spacetimes with odd dimensions}
For an odd number of spacetime dimensions with $D=2N+1$, the thermodynamic quantities become
\begin{eqnarray}
&& E=\frac{m}{\Xi} \left( \sum_{i=1}^N \frac{1}{\Xi_i}-\frac{1}{2}\right), \quad J_i=\frac{m a_i}{\Xi_i \Xi},\nonumber\\
&&S=\frac{\pi}{r_{\mathrm{H}}}\prod_{i=1}^N \frac{r_{\mathrm{H}}^2+a_{i}^2}{\Xi_i},\nonumber\\
&& T=\frac{r_{\mathrm{H}}\left(1+g^2 r_{\mathrm{H}}^2\right)}{2 \pi} \sum_{i=1}^N \frac{1}{r_{\mathrm{H}}^2+a_i^2}-\frac{1}{2 \pi r_{\mathrm{H}}}, \nonumber\\
&&\Omega_i=\frac{\left(1+g^2 r_{\mathrm{H}}^2\right) a_i}{r_{\mathrm{H}}^2+a_i^2}, \,\,\, 
C = \pm \frac{\sqrt{\vert \Lambda \vert}}{\Omega_{D-2}}, \nonumber\\
&& \left(1+g^2 r_{\mathrm{H}}^2\right) \prod_{i=1}^N \left(r_{\mathrm{H}}^2+a_i^2\right)-2 m r_{\mathrm{H}}^2=0,
\end{eqnarray}
where the first law and the Smarr relation reads the same as (\ref{firstlaw}) and (\ref{Smarr}), and the Schwarzschild limit yields the same result as in the even-dimensional case. The effective volume for this odd-dimensional case differs from that of the even-dimensional, case which can be found to be
\begin{equation}
    V_{\text{eff}}=\frac{m \Omega_{D-2}}{\Lambda \Xi}\left(-\sum_{i=1}^N \frac{1}{\Xi_i}-\frac{D-3}{2}+\frac{D-2}{1+g^2 r_{\mathrm{H}}^2}\right). 
\end{equation}
For asymptotically flat rotating black holes, one obtains the same limit (\ref{28}). 

\section{Thermodynamics of Einstein-Gauss-Bonnet Gravity}
Let us consider the action 
\begin{equation}
I = \int d^D x\sqrt{-g} \left(\frac{1}{4 \Omega_{D-2}}\left(R+\lambda\mathcal{L}_{GB}\right )\right), \label{action1}
\end{equation}
with the Gauss-Bonnet (GB) scalar given as $\mathcal{L}_{\text{GB}}:=R^{\mu\nu\rho\sigma}R_{\mu\nu\rho\sigma}-4 R^{\mu\nu}R_{\mu\nu}+R^2$, and $\lambda$ is a coupling constant of $M^2$ dimension. This theory has the same particle content as general relativity and arises as an effective model in string theory. In $D=4$, the GB part is topological and does not contribute to the field equations, but it does play a role in the thermodynamics of the black holes. Unfortunately, no rotating solution is known, but the spherically symmetric nonrotating solution was given in \cite{Boulwere1}, and the line element reads as $ds^2 = g_{tt} dt^2 + g_{rr} dr^2 + r^{2} d\Omega_{D-2}$ with
\begin{eqnarray}
-g_{tt} = g_{rr}^{-1} = &&1 + \left(\frac{r^2}{2 \lambda (D-3) (D-4)}\right)\\
&&\times\left (1 - \sqrt{1+\frac{8 m \lambda (D-3) (D-4)}{r^{D-1}}}\right),\nonumber
\end{eqnarray}
which is asymptotically flat.  One should note that the $D \rightarrow 4$ limit exists at the level of the action, but does not exist at the level of the solution \cite{Tahsin1, Tahsin2} since the GB term is a topological invariant in that dimension. Therefore for $D=4$, the Schwarzschild metric is the unique spherically symmetric solution of which some thermodynamical coordinates receive corrections from the topological term as can be seen below. 
The location of the event horizon is the largest root of the polynomial
\begin{equation}
2 m = r_{\mathrm{H}}^{D-3} + (D-3) (D-4) \lambda r_{\mathrm{H}}^{D-5}.
\end{equation}
The temperature and the entropy are given as
\begin{eqnarray}
T&=&\frac{(D-3) r_{\mathrm{H}}^2 + \lambda (D-3) (D-4) (D-5)  }{4 \pi r_{\mathrm{H}}^3 + 8 \pi r_{\mathrm{H}} \lambda (D-3) (D-4)}, \nonumber\\
S&=&\pi r_{\mathrm{H}}^{D-2} \left(1+\frac{2\lambda (D-2) (D-3)}{r_{\mathrm{H}}^2} \right),
\end{eqnarray}
respectively.  Observe that, for $D=4$, the entropy of the Schwarzschild black hole receives a correction from the topological GB term. The energy is given by \cite{Deser2}
\begin{equation}
E =\frac{(D-2)}{2}m.
\end{equation}
These thermodynamic quantities satisfy the following Smarr relation and the first law:
\begin{equation}
(D-3) E =(D-2) T S + 2 (D-3) (D-4) \lambda \Phi \label{SmarrGB}
\end{equation}
and 
\begin{equation}
\delta E = T \delta S + (D-3)(D-4) \Phi \delta \lambda,
\end{equation}
where the potential corresponding to the Gauss-Bonnet coupling constant is
\begin{equation}
\Phi=\frac{(D-2) r_{\mathrm{H}}^{D-5}}{4(D-4)} \left(\frac{-r_{\mathrm{H}}^2 (D-2) + 2\lambda (D-3)(D-4)}{r_{\mathrm{H}}^2 + 2 \lambda (D-3) (D-4)}\right),
\end{equation}
which is derived explicitly in Appendixes B and C by using the method developed in \cite{Mann1}. For $D=4$, this potential is divergent, but the correct limit in the Smarr formula (\ref{SmarrGB}) yields a finite contribution that reads 
\begin{equation}
E = 2 TS -  \frac{2 \lambda}{r_{\mathrm{H}}},
\end{equation}
which was first derived in \cite{Liberati1} and was called the topological work. 

\section{Conclusions} 
Understanding the macroscopic and microscopic thermodynamics of the gravitational field, especially in the case of strong gravitational fields of black holes will probably get us close to the quantum regime of gravity. Several decades of works along this direction have shown that the stationary gravitational fields that solve Einstein's theory with a cosmological constant obey the laws of thermodynamics in close analogy with matter fields, so close that one can talk about black hole chemistry, just like the chemistry of fluids with the usual volume, pressure, and other thermodynamic coordinates. But there is a major difference between the thermodynamics of matter and the gravitational field: in the former, we have a consistent microscopic theory, and we can leave the regime of thermodynamics that captures the gross features of a many-body system behind and work in the microscopic theory when high energy processes are involved, while in the latter we only have proposed effective theories, instead of a consistent microscopic theory, in which the Einstein-Hilbert action is augmented with higher powers of curvature. With the addition of these higher powers, one needs to define effective thermodynamics. These higher powers bring in new scales to the theory and modify the beautiful scaling laws that lead to the Gibbs-Duhem-Smarr formula, a relation that ties the thermodynamics coordinates that describe the gross features of the gravitational field. This relation must be consistent with the first law of thermodynamics. This requirement forces one to carefully define the conserved charges and thermodynamical coordinates of the gravitational field. Recently \cite{Hajian1} a generic procedure was outlined on how consistent thermodynamics can be defined in the presence of many dimensionful coupling constants by introducing auxiliary fields. In the current work, we have used those ideas plus the conserved charge construction in a generic gravity theory based on the background, or asymptotic Killing symmetries instead of the Komar charges, which is difficult to write for a generic theory. The versatility of this description is that it applies to any gravity theory that has a maximally symmetric vacuum solution. We applied the techniques to the rotating solutions of the $D$-dimensional cosmological Einstein theory and the spherically symmetric solutions of the Einstein-Gauss-Bonnet theory. The refined first law and the refined Smarr relations are satisfied. As a by-product of these refined laws, one gets an effective volume of black holes in generic $D$ dimensions for both asymptotically anti\textendash{}de Sitter and flat black holes. Also, the volume and the surface area of the black holes satisfy the reverse isoperimetric inequality, which becomes an equality for the Schwarzschild black hole.

\clearpage
\newpage
\onecolumngrid
\section{Appendix A: Thermodynamic Quantities for Kerr-Ads Spacetimes}
For an action of the form
$$
I=\frac{1}{16 \pi} \int d^D x \sqrt{-g}\left[R-2\Lambda \right],
$$
with a metric in generalized Boyer-Lindquist coordinates
\begin{eqnarray}
    ds^2 &=& -W \left(1+g^2 r^2\right) dt^2 + \frac{2 m}{U} \left(W dt - \sum_{i=1}^N \frac{a_i \mu_i^2}{\Xi_i}d\phi_i\right)^2 + \sum_{i=1}^N \frac{r^2 + a_i^2}{\Xi_i}\left(\mu_i^2 d\phi_i^2 + d\mu_i^2 \right) \nonumber\\
    &&+\frac{U}{V - 2m} dr^2 - \frac{g^2}{W \left(1 + g^2 r^2 \right)} \left(\sum_{i=1}^N \frac{r^2 + a_i^2}{\Xi_i} \mu_i d\mu_i + \epsilon r^2 \nu d\nu \right)^2 + \epsilon r^2 d\nu^2,
\end{eqnarray}
where
\begin{eqnarray}
    &&W := \sum_{i=1}^N \frac{\mu_i^2}{\Xi} + \epsilon \nu^2, \quad V := r^{\epsilon-2} \left(1 + g^2 r^2 \right) \prod_{i=1}^N \left(r^2 + a_i^2 \right), \quad U := \frac{V}{1 + g^2 r^2} \left(1 - \sum_{i=1}^N \frac{a_i^2 \mu_i^2}{r^2 + a_i^2} \right),
\end{eqnarray}
with the additional constraint
\begin{equation}
    \sum_{i=1}^N \mu_i^2 + \epsilon \nu^2 =1,
\end{equation}
the thermodynamic quantities are derived in \cite{Cvetic1}. Here, $D=2N+2$ and $\epsilon=1$ for an even number of dimensions, and $D=2N+1$ and $\epsilon=0$ for an odd number of dimensions.
However, in this current work, an action of the form
\begin{equation}
I=\frac{1}{4\Omega_{D-2}}\int d^Dx\sqrt{-g}\left(R-2\Lambda\right),
\end{equation}
is being used, and as a result, some of these thermodynamic quantities should be rescaled due to the differences between the normalization constants. In the first part of this section, an even number of dimensions with $D=2N+2$ is considered. The temperature, the angular velocities, and the equation defining the event horizon are directly related to the metric, which are independent of normalization constants. Hence, they remain the same. Nevertheless, the entropy is directly related to the form of the action and should be rescaled; one gets
$$
\begin{aligned}
& S=\pi\prod_{i=1}^N \frac{r_{\mathrm{H}}^2+a_{i}^2}{\Xi_i}, \quad T=\frac{r_{\mathrm{H}}\left(1+g^2 r_{\mathrm{H}}^2\right)}{2 \pi} \sum_{i=1}^{N} \frac{1}{r_{\mathrm{H}}^2+a_i^2}-\frac{1-g^2 r_{\mathrm{H}}^2}{4 \pi r_{\mathrm{H}}},  \\
& \Omega_i=\frac{\left(1+g^2 r_{\mathrm{H}}^2\right) a_i}{r_{\mathrm{H}}^2+a_i^2}, \quad 2 m=\frac{1}{r_{\mathrm{H}}}\left(1+g^2 r_{\mathrm{H}}^2\right) \prod_{i=1}^{N} \left(r_{\mathrm{H}}^2+a_i^2\right). \\
\end{aligned}
$$
The energy and angular momenta are derived in \cite{Deser3}, which read as
\begin{equation}
E=\frac{m}{\Xi} \sum_{i=1}^N \frac{1}{\Xi_i}, \quad J_i=\frac{m a_i}{\Xi_i \Xi}.
\end{equation}
The cosmological charge and cosmological potential are derived in \cite{Hajian2} and, after rescaling the cosmological charge term due to the difference between the normalization constants of two actions, read as
\begin{equation}
C=\pm\frac{\sqrt{\vert\Lambda\vert}}{\Omega_{D-2}} \quad \text{and} \quad \Theta_{\mathrm{H}}=\pm \sqrt{\vert\Lambda\vert}V_{\text{eff}}.
\end{equation}
After inserting these quantities into the Smarr relation, one gets
\begin{eqnarray}
&&\left(D-3\right) E = \left(D-2\right) \left(T S + \sum_{i=1}^N \Omega_i J_i \right)- \Theta_{\mathrm{H}} C,\nonumber\\
&&\left(D-3\right)\left(\frac{m}{\Xi} \sum_{i=1}^N \frac{1}{\Xi_i}\right) = \left(D-2\right) \left(\left(\frac{r_{\mathrm{H}}\left(1+g^2 r_{\mathrm{H}}^2\right)}{2 \pi} \sum_{i=1}^{N} \frac{1}{r_{\mathrm{H}}^2+a_i^2}-\frac{1-g^2 r_{\mathrm{H}}^2}{4 \pi r_{\mathrm{H}}}\right)\left(\pi\prod_{i=1}^N \frac{r_{\mathrm{H}}^2+a_{i}^2}{\Xi_i}\right)\right.\nonumber\\
&&\left.+\left( \sum_{i=1}^N \left(\frac{\left(1+g^2 r_{\mathrm{H}}^2\right) a_i}{r_{\mathrm{H}}^2+a_i^2}\right)\left(\frac{m a_i}{\Xi_i \Xi}\right)\right)\right)-\left(\frac{\sqrt{\vert\Lambda\vert}}{\Omega_{D-2}}\right)\left(\sqrt{\vert\Lambda\vert}V_{\text{eff}}\right),\nonumber\\
&&\left(D-3\right)\left(\frac{m}{\Xi} \sum_{i=1}^N \frac{1}{\Xi_i}\right)-\left(D-2\right)\left(\frac{m}{\Xi}\sum_{i=1}^N\left(\frac{\left(1+g^2 r_{\mathrm{H}}^2\right)a_i^2}{\left(r_{\mathrm{H}}^2+a_i^2\right)\Xi_i}\right)\right)=\left(D-2\right)\nonumber\\
&&\times\left(\left(\frac{r_{\mathrm{H}}\left(1+g^2 r_{\mathrm{H}}^2\right)}{2 \pi} \sum_{i=1}^{N} \frac{1}{r_{\mathrm{H}}^2+a_i^2}-\frac{1-g^2 r_{\mathrm{H}}^2}{4 \pi r_{\mathrm{H}}}\right)\left(\pi\prod_{i=1}^N \frac{r_{\mathrm{H}}^2+a_{i}^2}{\Xi_i}\right)\right)-\left(\frac{\sqrt{\vert\Lambda\vert}}{\Omega_{D-2}}\right)\left(\sqrt{\vert\Lambda\vert}V_{\text{eff}}\right),\nonumber\\
&&\left(D-3\right)\left(\frac{m}{\Xi} \sum_{i=1}^N \frac{1}{\Xi_i}\right)-\left(D-2\right)\left(\frac{m \left(1+g^2 r_{\mathrm{H}}^2\right)}{\Xi}\sum_{i=1}^N\left(\frac{a_i^2}{\left(r_{\mathrm{H}}^2+a_i^2\right)\Xi_i}\right)\right)=\left(D-2\right)\nonumber\\
&&\times\left(\frac{1}{\Xi}\left(\frac{r_{\mathrm{H}}\left(1+g^2 r_{\mathrm{H}}^2\right)}{2} \sum_{i=1}^{N} \frac{1}{r_{\mathrm{H}}^2+a_i^2}-\frac{1-g^2 r_{\mathrm{H}}^2}{4 r_{\mathrm{H}}}\right)\left(\prod_{i=1}^N r_{\mathrm{H}}^2+a_{i}^2\right)\right)-\left(\frac{\sqrt{\vert\Lambda\vert}}{\Omega_{D-2}}\right)\left(\sqrt{\vert\Lambda\vert}V_{\text{eff}}\right),\nonumber\\
&&\left(D-3\right)\left(m \sum_{i=1}^N \frac{1}{\Xi_i}\right)-\left(D-2\right)\left(m \left(1+g^2 r_{\mathrm{H}}^2\right)\sum_{i=1}^N\left(\frac{a_i^2}{\left(r_{\mathrm{H}}^2+a_i^2\right)\Xi_i}\right)\right)=\left(D-2\right)\nonumber\\
&&\times\left(\left(\frac{r_{\mathrm{H}}\left(1+g^2 r_{\mathrm{H}}^2\right)}{2} \sum_{i=1}^{N} \frac{1}{r_{\mathrm{H}}^2+a_i^2}-\frac{1-g^2 r_{\mathrm{H}}^2}{4 r_{\mathrm{H}}}\right)\left(\prod_{i=1}^N r_{\mathrm{H}}^2+a_{i}^2\right)\right)-\Xi\left(\frac{\sqrt{\vert\Lambda\vert}}{\Omega_{D-2}}\right)\left(\sqrt{\vert\Lambda\vert}V_{\text{eff}}\right),\nonumber\\
&&\left(D-3\right)\left(m \sum_{i=1}^N \frac{1}{\Xi_i}\right)-\left(D-2\right)\left(m \left(1+g^2 r_{\mathrm{H}}^2\right)\sum_{i=1}^N\left(\frac{a_i^2}{\left(r_{\mathrm{H}}^2+a_i^2\right)\Xi_i}\right)\right)=\left(D-2\right)\nonumber\\
&&\times\left(\left(\frac{r_{\mathrm{H}}\left(1+g^2 r_{\mathrm{H}}^2\right)}{2} \sum_{i=1}^{N} \frac{1}{r_{\mathrm{H}}^2+a_i^2}-\frac{1-g^2 r_{\mathrm{H}}^2}{4 r_{\mathrm{H}}}\right)\left(\frac{2 m r_{\mathrm{H}}}{1 + g^2 r_{\mathrm{H}}^2}\right)\right)-\Xi\left(\frac{\sqrt{\vert\Lambda\vert}}{\Omega_{D-2}}\right)\left(\sqrt{\vert\Lambda\vert}V_{\text{eff}}\right),\nonumber\\
&&\left(D-3\right)\left(\sum_{i=1}^N \frac{1}{\Xi_i}\right)-\left(D-2\right)\left(\left(1+g^2 r_{\mathrm{H}}^2\right)\sum_{i=1}^N\left(\frac{a_i^2}{\left(r_{\mathrm{H}}^2+a_i^2\right)\Xi_i}\right)\right)=\left(D-2\right)\nonumber\\
&&\times\left( r_{\mathrm{H}}^2\sum_{i=1}^{N} \frac{1}{r_{\mathrm{H}}^2+a_i^2}-\frac{1-g^2 r_{\mathrm{H}}^2}{2 \left(1+g^2 r_{\mathrm{H}}^2\right)}\right)-\frac{\Xi}{m}\left(\frac{\sqrt{\vert\Lambda\vert}}{\Omega_{D-2}}\right)\left(\sqrt{\vert\Lambda\vert}V_{\text{eff}}\right),\nonumber\\
&&\left(D-3\right)\left(\sum_{i=1}^N \frac{1}{\Xi_i}\right)-\left(D-2\right)\left(r_{\mathrm{H}}^2\sum_{i=1}^{N} \frac{1}{r_{\mathrm{H}}^2+a_i^2}+\left(1+g^2 r_{\mathrm{H}}^2\right)\sum_{i=1}^N\left(\frac{a_i^2}{\left(r_{\mathrm{H}}^2+a_i^2\right)\Xi_i}\right)\right)=\left(D-2\right)\nonumber\\
&&\times\left(-\frac{1-g^2 r_{\mathrm{H}}^2}{2 \left(1+g^2 r_{\mathrm{H}}^2\right)}\right)-\frac{\Xi}{m}\left(\frac{\sqrt{\vert\Lambda\vert}}{\Omega_{D-2}}\right)\left(\sqrt{\vert\Lambda\vert}V_{\text{eff}}\right),\nonumber\\
&&\left(D-3\right)\left(\sum_{i=1}^N \frac{1}{\Xi_i}\right)-\left(D-2\right)\left(\sum_{i=1}^N \frac{1}{\Xi_i}\right)=\left(D-2\right)\nonumber\\
&&\times\left(-\frac{1-g^2 r_{\mathrm{H}}^2}{2 \left(1+g^2 r_{\mathrm{H}}^2\right)}\right)-\frac{\Xi}{m}\left(\frac{\sqrt{\vert\Lambda\vert}}{\Omega_{D-2}}\right)\left(\sqrt{\vert\Lambda\vert}V_{\text{eff}}\right),\nonumber\\
&&\left(\sum_{i=1}^N \frac{1}{\Xi_i}\right)=\left(D-2\right)\left(\frac{1-g^2 r_{\mathrm{H}}^2}{2 \left(1+g^2 r_{\mathrm{H}}^2\right)}\right)+\frac{\Xi}{m}\left(\frac{\sqrt{\vert\Lambda\vert}}{\Omega_{D-2}}\right)\left(\sqrt{\vert\Lambda\vert}V_{\text{eff}}\right),\nonumber\\
&&\frac{m}{\Xi}\left(\left(\sum_{i=1}^N \frac{1}{\Xi_i}\right)-\left(D-2\right)\left(\frac{1-g^2 r_{\mathrm{H}}^2}{2 \left(1+g^2 r_{\mathrm{H}}^2\right)}\right)\right)=\left(\frac{-\Lambda}{\Omega_{D-2}}\right)V_{\text{eff}},\nonumber\\
&&V_{\text{eff}}=\frac{m \Omega_{D-2}}{\Lambda \Xi}\left(-\sum_{i=1}^N \frac{1}{\Xi_i}+\left(\frac{D-2}{2}\right)\left(\frac{1-g^2 r_{\mathrm{H}}^2}{1+g^2 r_{\mathrm{H}}^2}\right)\right). 
\end{eqnarray}
One can derive the thermodynamic quantities for odd dimensions by using the same approach and get
\begin{eqnarray}
&&\left(D-3\right) E = \left(D-2\right) \left(T S + \sum_{i=1}^N \Omega_i J_i \right)- \Theta_{\mathrm{H}} C,\nonumber\\
&&\left(D-3\right)\left(\frac{m}{\Xi}\right)\left(\sum_{i=1}^N \frac{1}{\Xi_i} - \frac{1}{2}\right) = \left(D-2\right) \left(\left(\frac{r_{\mathrm{H}}\left(1+g^2 r_{\mathrm{H}}^2\right)}{2 \pi} \sum_{i=1}^{N} \frac{1}{r_{\mathrm{H}}^2+a_i^2}-\frac{1}{2 \pi r_{\mathrm{H}}}\right)\left(\frac{\pi}{r_{\mathrm{H}}}\prod_{i=1}^N \frac{r_{\mathrm{H}}^2+a_{i}^2}{\Xi_i}\right)\right.\nonumber\\
&&\left.+\left( \sum_{i=1}^N \left(\frac{\left(1+g^2 r_{\mathrm{H}}^2\right) a_i}{r_{\mathrm{H}}^2+a_i^2}\right)\left(\frac{m a_i}{\Xi_i \Xi}\right)\right)\right)-\left(\frac{\sqrt{\vert\Lambda\vert}}{\Omega_{D-2}}\right)\left(\sqrt{\vert\Lambda\vert}V_{\text{eff}}\right),\nonumber\\
&&\left(D-3\right)\left(\frac{m}{\Xi} \sum_{i=1}^N \frac{1}{\Xi_i}\right)-\left(D-3\right)\left(\frac{m}{2\Xi}\right)-\left(D-2\right)\left(\frac{m}{\Xi}\sum_{i=1}^N\left(\frac{\left(1+g^2 r_{\mathrm{H}}^2\right)a_i^2}{\left(r_{\mathrm{H}}^2+a_i^2\right)\Xi_i}\right)\right)=\left(D-2\right)\nonumber\\
&&\times\left(\left(\frac{r_{\mathrm{H}}\left(1+g^2 r_{\mathrm{H}}^2\right)}{2 \pi} \sum_{i=1}^{N} \frac{1}{r_{\mathrm{H}}^2+a_i^2}-\frac{1}{2 \pi r_{\mathrm{H}}}\right)\left(\frac{\pi}{r_{\mathrm{H}}}\prod_{i=1}^N \frac{r_{\mathrm{H}}^2+a_{i}^2}{\Xi_i}\right)\right)-\left(\frac{\sqrt{\vert\Lambda\vert}}{\Omega_{D-2}}\right)\left(\sqrt{\vert\Lambda\vert}V_{\text{eff}}\right),\nonumber\\
&&\left(D-3\right)\left(\frac{m}{\Xi} \sum_{i=1}^N \frac{1}{\Xi_i}\right)-\left(D-3\right)\left(\frac{m}{2\Xi}\right)-\left(D-2\right)\left(\frac{m \left(1+g^2 r_{\mathrm{H}}^2\right)}{\Xi}\sum_{i=1}^N\left(\frac{a_i^2}{\left(r_{\mathrm{H}}^2+a_i^2\right)\Xi_i}\right)\right)=\left(D-2\right)\nonumber\\
&&\times\left(\frac{1}{\Xi}\left(\frac{\left(1+g^2 r_{\mathrm{H}}^2\right)}{2} \sum_{i=1}^{N} \frac{1}{r_{\mathrm{H}}^2+a_i^2}-\frac{1}{2 r_{\mathrm{H}}^2}\right)\left(\prod_{i=1}^N r_{\mathrm{H}}^2+a_{i}^2\right)\right)-\left(\frac{\sqrt{\vert\Lambda\vert}}{\Omega_{D-2}}\right)\left(\sqrt{\vert\Lambda\vert}V_{\text{eff}}\right),\nonumber\\
&&\left(D-3\right)\left(m \sum_{i=1}^N \frac{1}{\Xi_i}\right)-\left(D-3\right)\left(\frac{m}{2}\right)-\left(D-2\right)\left(m \left(1+g^2 r_{\mathrm{H}}^2\right)\sum_{i=1}^N\left(\frac{a_i^2}{\left(r_{\mathrm{H}}^2+a_i^2\right)\Xi_i}\right)\right)=\left(D-2\right)\nonumber\\
&&\times\left(\left(\frac{\left(1+g^2 r_{\mathrm{H}}^2\right)}{2} \sum_{i=1}^{N} \frac{1}{r_{\mathrm{H}}^2+a_i^2}-\frac{1}{2 r_{\mathrm{H}}^2}\right)\left(\prod_{i=1}^N r_{\mathrm{H}}^2+a_{i}^2\right)\right)-\Xi\left(\frac{\sqrt{\vert\Lambda\vert}}{\Omega_{D-2}}\right)\left(\sqrt{\vert\Lambda\vert}V_{\text{eff}}\right),\nonumber\\
&&\left(D-3\right)\left(m \sum_{i=1}^N \frac{1}{\Xi_i}\right)-\left(D-3\right)\left(\frac{m}{2}\right)-\left(D-2\right)\left(m \left(1+g^2 r_{\mathrm{H}}^2\right)\sum_{i=1}^N\left(\frac{a_i^2}{\left(r_{\mathrm{H}}^2+a_i^2\right)\Xi_i}\right)\right)=\left(D-2\right)\nonumber\\
&&\times\left(\left(\frac{\left(1+g^2 r_{\mathrm{H}}^2\right)}{2} \sum_{i=1}^{N} \frac{1}{r_{\mathrm{H}}^2+a_i^2}-\frac{1}{2 r_{\mathrm{H}}^2}\right)\left(\frac{2 m r_{\mathrm{H}}^2}{1 + g^2 r_{\mathrm{H}}^2}\right)\right)-\Xi\left(\frac{\sqrt{\vert\Lambda\vert}}{\Omega_{D-2}}\right)\left(\sqrt{\vert\Lambda\vert}V_{\text{eff}}\right),\nonumber\\
&&\left(D-3\right)\left(\sum_{i=1}^N \frac{1}{\Xi_i}\right)-\left(\frac{D-3}{2}\right)-\left(D-2\right)\left(\left(1+g^2 r_{\mathrm{H}}^2\right)\sum_{i=1}^N\left(\frac{a_i^2}{\left(r_{\mathrm{H}}^2+a_i^2\right)\Xi_i}\right)\right)=\left(D-2\right)\nonumber\\
&&\times\left( r_{\mathrm{H}}^2\sum_{i=1}^{N} \frac{1}{r_{\mathrm{H}}^2+a_i^2}-\frac{1}{1+g^2 r_{\mathrm{H}}^2}\right)-\frac{\Xi}{m}\left(\frac{\sqrt{\vert\Lambda\vert}}{\Omega_{D-2}}\right)\left(\sqrt{\vert\Lambda\vert}V_{\text{eff}}\right),\nonumber\\
&&\left(D-3\right)\left(\sum_{i=1}^N \frac{1}{\Xi_i}\right)-\left(\frac{D-3}{2}\right)-\left(D-2\right)\left(r_{\mathrm{H}}^2\sum_{i=1}^{N} \frac{1}{r_{\mathrm{H}}^2+a_i^2}+\left(1+g^2 r_{\mathrm{H}}^2\right)\sum_{i=1}^N\left(\frac{a_i^2}{\left(r_{\mathrm{H}}^2+a_i^2\right)\Xi_i}\right)\right)=\left(D-2\right)\nonumber\\
&&\times\left(-\frac{1}{1+g^2 r_{\mathrm{H}}^2}\right)-\frac{\Xi}{m}\left(\frac{\sqrt{\vert\Lambda\vert}}{\Omega_{D-2}}\right)\left(\sqrt{\vert\Lambda\vert}V_{\text{eff}}\right),\nonumber\\
&&\left(D-3\right)\left(\sum_{i=1}^N \frac{1}{\Xi_i}\right)-\left(\frac{D-3}{2}\right)-\left(D-2\right)\left(\sum_{i=1}^N \frac{1}{\Xi_i}\right)=\left(D-2\right)\nonumber\\
&&\times\left(-\frac{1}{1+g^2 r_{\mathrm{H}}^2}\right)-\frac{\Xi}{m}\left(\frac{\sqrt{\vert\Lambda\vert}}{\Omega_{D-2}}\right)\left(\sqrt{\vert\Lambda\vert}V_{\text{eff}}\right),\nonumber\\
&&\left(\sum_{i=1}^N \frac{1}{\Xi_i}\right)+\left(\frac{D-3}{2}\right)=\left(D-2\right)\left(\frac{1}{1+g^2 r_{\mathrm{H}}^2}\right)+\frac{\Xi}{m}\left(\frac{\sqrt{\vert\Lambda\vert}}{\Omega_{D-2}}\right)\left(\sqrt{\vert\Lambda\vert}V_{\text{eff}}\right),\nonumber\\
&&\frac{m}{\Xi}\left(\left(\sum_{i=1}^N \frac{1}{\Xi_i}\right)+\left(\frac{D-3}{2}\right)-\left(D-2\right)\left(\frac{1}{1+g^2 r_{\mathrm{H}}^2}\right)\right)=\left(\frac{-\Lambda}{\Omega_{D-2}}\right)V_{\text{eff}},\nonumber\\
&&V_{\text{eff}}=\frac{m \Omega_{D-2}}{\Lambda \Xi}\left(-\sum_{i=1}^N \frac{1}{\Xi_i}-\left(\frac{D-3}{2}\right)+\left(D-2\right)\left(\frac{1}{1+g^2 r_{\mathrm{H}}^2}\right)\right). 
\end{eqnarray}
\section{Appendix B: Derivation of the Charge Term  for Einstein-Gauss-Bonnet Gravity in Generic $D$ Dimensions}
For an action of the form
\begin{equation}
I = \int d^D x \sqrt{-g} \frac{1}{16 \pi G} \left(R - 2 \Lambda + \frac{\lambda_{\text{GB}} \mathcal{L}}{(D-3)(D-4)}\right),
\end{equation}
with a metric of the form
\begin{equation}
ds^2 = - f(r) dt^2 + \frac{dr^2}{f(r)} + r^2 d\Omega^2,
\end{equation}
where
\begin{equation}
f(r) = 1 + \frac{r^2}{2 \lambda_{\text{GB}}} - \frac{r^{2-D/2}\sqrt{r^D+4 \lambda_{\text{GB}}\left(r \omega_{D-3}+r^D/L^2\right)}}{2 \lambda_{\text{GB}}},
\end{equation}
\begin{equation}
\Lambda=\frac{(D-1)(D-2)}{2L^2},
\end{equation}
and
\begin{equation}
\omega_{D-3}=\frac{16 \pi G m}{(D-2)\Omega_{D-2}},
\end{equation}
the thermodynamic quantities are calculated in \cite{Mann1}. Another important function defined in \cite{Mann1} is the pure dS vacuum of the theory that reads as
\begin{equation}
f_0(r) = 1+r^2\left(\frac{1-\sqrt{1+\frac{4 \lambda_{\text{GB}}}{L^2}}}{2 \lambda_{\text{GB}}}\right).
\end{equation}
In the asymptotically flat limit, $\Lambda \rightarrow 0$, the case that is being investigated in our current work, these terms become
\begin{equation}
f(r) = 1 + \frac{r^2}{2 \lambda_{\text{GB}}} - \frac{\sqrt{r^4+4\lambda_{\text{GB}}\omega_{D-3}r^{5-D}}}{2\lambda_{\text{GB}}}
\end{equation}
and
\begin{equation}
f_0(r) = 1.
\end{equation}
\subsection{Energy}
The energy term is given in \cite{Mann1} as
\begin{equation}
\begin{aligned}
E=& \frac{(D-2) \Omega_{D-2} r_c^{D-5}}{24 \pi G}\left[\sqrt{f\left(r_c\right)}\left(-3 r_c^2+2 \lambda_{\mathrm{GB}}\left(f\left(r_c\right)-3\right)\right)\right. \\
& \left.-\sqrt{f_0\left(r_c\right)}\left(-3 r_c^2+2 \lambda_{\mathrm{GB}}\left(f_0\left(r_c\right)-3\right)\right)\right],
\end{aligned}
\end{equation}
where $r_c$ gives the location of the cavity. 
In the asymptotically flat limit, $\Lambda \rightarrow 0$ and $r_c \rightarrow \infty $, one gets
\begin{equation}
E=m.
\end{equation}
\subsection{Entropy}
The entropy term is given in \cite{Mann1} as
\begin{equation}
S = \frac{\Omega_{D-2} r_{\mathrm{H}}^{D-2}}{4G} + \frac{(D-2) \lambda_{\text{GB}} \Omega_{D-2}r_{\mathrm{H}}^{D-4}}{2 G (D-4)},
\end{equation}
where
\begin{equation}
\lambda_{\text{GB}} + r_{\mathrm{H}}^{2} = r_{\mathrm{H}}^{5-D} \omega_{D-3}.
\end{equation}
\subsection{Temperature}
The horizon temperature is
\begin{equation}
T = \frac{1}{4\pi}\left(\frac{(D-3) r_{\mathrm{H}}^2 + (D-5) \lambda_{\text{GB}}}{r_{\mathrm{H}}^3 + 2 \lambda_{\text{GB}} r_{\mathrm{H}}}\right).
\end{equation}
\subsection{Charge}
In the asymptotically flat limit, the Smarr relation is given in \cite{Mann1} as
\begin{equation}
(D-3) E = (D-2) TS + 2 \Phi_{\text{GB}} \lambda_{\text{GB}}.
\end{equation}
Therefore, the charge term can be found as
\begin{equation}
\Phi_{\text{GB}} = \frac{(D-3)E - (D-2)TS}{2\lambda_{\text{GB}}}.
\end{equation}
It is known that 
\begin{equation}
E = m = \frac{(D-2) \Omega_{D-2} \omega_{D-3}}{16 \pi G } = \frac{(D-2) \Omega_{D-2} \left(r_{\mathrm{H}}^{D-5} \lambda_{\text{GB}} + r_{\mathrm{H}}^{D-3}\right)}{16 \pi G}.
\end{equation}
When the corresponding terms are inserted, one gets
\begin{eqnarray}
\Phi_{\text{GB}} &=& \frac{1}{2 \lambda_{\text{GB}}}\left[(D-3) \left(\frac{(D-2) \Omega_{D-2} \left(r_{\mathrm{H}}^{D-5} \lambda_{\text{GB}} + r_{\mathrm{H}}^{D-3}\right)}{16 \pi G}\right)\right.\nonumber\\
&&\left.-(D-2)\left(\frac{1}{4\pi}\left(\frac{(D-3) r_{\mathrm{H}}^2 + (D-5) \lambda_{\text{GB}}}{r_{\mathrm{H}}^3 + 2 \lambda_{\text{GB}} r_{\mathrm{H}}}\right)\right)\left(\frac{\Omega_{D-2} r_{\mathrm{H}}^{D-2}}{4G} + \frac{(D-2) \lambda_{\text{GB}} \Omega_{D-2}r_{\mathrm{H}}^{D-4}}{2 G (D-4)}\right)\right]\nonumber\\
&=&\frac{(D-2) \Omega_{D-2} r_{\mathrm{H}}^{D-5}}{32 \pi G \lambda_{\text{GB}}}\left[\left((D-3) \left(\lambda_{\text{GB}}+r_{\mathrm{H}}^2\right)\right)\right.\nonumber\\
&&\left.-\left(\frac{(D-3) r_{\mathrm{H}}^2 + (D-5) \lambda_{\text{GB}}}{r_{\mathrm{H}}^2 + 2 \lambda_{\text{GB}}}\right)\left(r_{\mathrm{H}}^2 +\frac{2(D-2) \lambda_{\text{GB}}}{(D-4)}\right)\right]\nonumber\\
&=&\frac{(D-2) \Omega_{D-2} r_{\mathrm{H}}^{D-5}}{32 \pi G \lambda_{\text{GB}}}\left[\left((D-3) \left(\lambda_{\text{GB}}+r_H^2\right)\right)\right.\nonumber\\
&&\left.-\left(\frac{(D-3) r_{\mathrm{H}}^2 + (D-5) \lambda_{\text{GB}}}{r_{\mathrm{H}}^2 + 2 \lambda_{\text{GB}}}\right)\left(\frac{r_{\mathrm{H}}^2 (D-4) + 2 (D-2) \lambda_{\text{GB}}}{(D-4)}\right)\right]\nonumber\\
&=&\frac{(D-2) \Omega_{D-2} r_{\mathrm{H}}^{D-5}}{32 \pi G \lambda_{\text{GB}} (D-4) \left(r_{\mathrm{H}}^2 + 2 \lambda_{\text{GB}}\right)}\left[\left((D-3) (D-4) \left(r_{\mathrm{H}}^2 + \lambda_{\text{GB}}\right)\left(r_{\mathrm{H}}^2 + 2\lambda_{\text{GB}}\right)\right)\right.\nonumber\\
&&\left.-\left((D-3) r_{\mathrm{H}}^2 + (D-5) \lambda_{\text{GB}}\right)\left((D-4) r_{\mathrm{H}}^2 + 2(D-2) \lambda_{\text{GB}}\right)\right]\nonumber\\
&=&\frac{(D-2) \Omega_{D-2} r_{\mathrm{H}}^{D-5}}{32 \pi G \lambda_{\text{GB}} (D-4) \left(r_{\mathrm{H}}^2 + 2 \lambda_{\text{GB}}\right)}\left[r_{\mathrm{H}}^4\left((D-3)(D-4)-(D-3) (D-4)\right)\right.\nonumber\\
&&\left.+\lambda_{\text{GB}} r_{\mathrm{H}}^2 \left(3 (D-3) (D-4) - 2 (D-2) (D-3) - (D-4) (D-5)\right)\right.\nonumber\\
&&\left.+\lambda_{\text{GB}}^2\left(2 (D-3) (D-4) - 2 (D-2) (D-5)\right)\right]\nonumber\\
&=&\frac{(D-2) \Omega_{D-2} r_{\mathrm{H}}^{D-5}}{32 \pi G \lambda_{\text{GB}} (D-4) \left(r_{\mathrm{H}}^2 + 2 \lambda_{\text{GB}}\right)}\left[\lambda_{\text{GB}} r_{\mathrm{H}}^2\left(3 D^2 - 21D + 36 - 2D^2 + 10D - 12 - D^2 + 9D - 20\right)\right.\nonumber\\
&&\left. +2\lambda_{\text{GB}}^2\left((D-3) (D-4) - (D-2) (D-5)\right)\right]\nonumber\\
&=&\frac{(D-2) \Omega_{D-2} r_{\mathrm{H}}^{D-5}}{32 \pi G (D-4) \left(r_{\mathrm{H}}^2 + 2 \lambda_{\text{GB}}\right)}\left[r_{\mathrm{H}}^2\left(-2D+4\right)+2\lambda_{\text{GB}}\left(D^2-7D+12-D^2+7D-10\right)\right]\nonumber\\
&=&\frac{(D-2) \Omega_{D-2} r_{\mathrm{H}}^{D-5}}{32 \pi G (D-4) \left(r_{\mathrm{H}}^2 + 2 \lambda_{\text{GB}}\right)}\left[-2r_{\mathrm{H}}^2\left(D-2\right)+4\lambda_{\text{GB}}\right]\nonumber\\
&=&\frac{(D-2) \Omega_{D-2} r_{\mathrm{H}}^{D-5}}{16 \pi G (D-4) \left(r_{\mathrm{H}}^2 + 2 \lambda_{\text{GB}}\right)}\left[-r_{\mathrm{H}}^2\left(D-2\right)+2\lambda_{\text{GB}}\right].
\end{eqnarray}
Because of differences between the normalization constants of two actions, the charge term associated with the action used in our work should be 
\begin{equation}
\Phi = \frac{16 \pi G}{4 \Omega_{D-2}}\Phi_{\text{GB}},
\end{equation}
and the coupling constant should be
\begin{equation}
\lambda_{\text{GB}} = (D-3) (D-4)\lambda.
\end{equation}
As a result, the corresponding charge term, the Smarr relation, and the first law become
\begin{eqnarray}
\Phi&=&\frac{(D-2) r_{\mathrm{H}}^{D-5}}{4 (D-4) \left(r_{\mathrm{H}}^2 + 2 \lambda (D-3) (D-4)\right)}\left[-r_{\mathrm{H}}^2\left(D-2\right)+2\lambda (D-3) (D-4)\right]\nonumber\\
&=&\frac{(D-2) r_{\mathrm{H}}^{D-5}}{4(D-4)} \left(\frac{-r_{\mathrm{H}}^2 (D-2) + 2\lambda (D-3)(D-4)}{r_{\mathrm{H}}^2 + 2 \lambda (D-3) (D-4)}\right),
\end{eqnarray}
\begin{equation}
(D-3) E = (D-2) TS + 2 (D-3) (D-4) \Phi \lambda,
\end{equation}
and
\begin{equation}
\delta E = T \delta S + (D-3) (D-4) \Phi \delta \lambda,
\end{equation}
respectively.
\section{Appendix C: Explicit Calculation of the Smarr Relation with Our Thermodynamical Quantities}
In our current work, the Smarr relation becomes
\begin{equation}
(D-3) E =(D-2) T S + 2 (D-3) (D-4) \lambda \tilde{\Phi},
\end{equation}
with the corresponding thermodynamic quantities
\begin{equation}
E = \frac{D-2}{2} m,
\end{equation}
\begin{equation}
T = \frac{(D-3) r_{\mathrm{H}}^2 + (D-3) (D-4) (D-5) \lambda}{4 \pi r_{\mathrm{H}}^3 + 8 \pi  r_{\mathrm{H}} \lambda (D-3) (D-4)},
\end{equation}
\begin{equation}
\tilde{S} = \pi r_{\mathrm{H}}^{D-2} + 2 (D-2) (D-3) \pi \lambda r_{\mathrm{H}}^{D-4},
\end{equation}
and 
\begin{equation}
2 M = r_{\mathrm{H}}^{D-3} + (D-3) (D-4) \lambda r_{\mathrm{H}}^{D-5}.
\end{equation}
When these terms are inserted into the Smarr relation, it becomes
\begin{eqnarray}
&&(D-3) \frac{(D-2)m}{2} = (D-2) \left(\frac{(D-3) r_{\mathrm{H}}^2 + (D-3) (D-4) (D-5) \lambda}{4 \pi r_{\mathrm{H}}^3 + 8 \pi  r_{\mathrm{H}} \lambda (D-3) (D-4)}\right) \left(\pi r_{\mathrm{H}}^{D-2} + 2 (D-2) (D-3) \pi \lambda r_{\mathrm{H}}^{D-4}\right)\nonumber\\
&&+ 2(D-3) (D-4) \lambda \left(\frac{(D-2) r_{\mathrm{H}}^{D-5}}{4(D-4)} \left(\frac{-r_{\mathrm{H}}^2 (D-2) + 2\lambda (D-3)(D-4)}{r_{\mathrm{H}}^2 + 2 \lambda (D-3) (D-4)}\right)\right)\nonumber,\\
&&\frac{M}{2} =  \left(\frac{r_{\mathrm{H}}^2 + (D-4) (D-5) \lambda}{4 r_{\mathrm{H}}^2 + 8 \lambda (D-3) (D-4)}\right)\left(r_{\mathrm{H}}^{D-3} + 2 (D-2) (D-3) \lambda r_{\mathrm{H}}^{D-5}\right)\nonumber\\
&&+ 2 \lambda \left(\frac{ r_{\mathrm{H}}^{D-5}}{4} \left(\frac{-r_{\mathrm{H}}^2 (D-2) + 2\lambda (D-3)(D-4)}{r_{\mathrm{H}}^2 + 2 \lambda (D-3) (D-4)}\right)\right)\nonumber,\\
&&\frac{\left(r_{\mathrm{H}}^{D-3} + (D-3) (D-4) \lambda r_{\mathrm{H}}^{D-5}\right)}{4} =  \left(\frac{r_{\mathrm{H}}^2 + (D-4) (D-5) \lambda}{4 r_{\mathrm{H}}^2 + 8 \lambda (D-3) (D-4)}\right)\left(r_{\mathrm{H}}^{D-3} + 2 (D-2) (D-3) \lambda r_{\mathrm{H}}^{D-5}\right)\nonumber\\
&&+ 2 \lambda \left(\frac{ r_{\mathrm{H}}^{D-5}}{4} \left(\frac{-r_{\mathrm{H}}^2 (D-2) + 2\lambda (D-3)(D-4)}{r_{\mathrm{H}}^2 + 2 \lambda (D-3) (D-4)}\right)\right)\nonumber,\\
&&\frac{\left(r_{\mathrm{H}}^2 + (D-3) (D-4) \lambda \right)}{4} =  \left(\frac{r_{\mathrm{H}}^2 + (D-4) (D-5) \lambda}{4 r_{\mathrm{H}}^2 + 8 \lambda (D-3) (D-4)}\right)\left(r_{\mathrm{H}}^2 + 2 (D-2) (D-3) \lambda \right)\nonumber\\
&&+ 2 \lambda \left(\frac{1 }{4} \left(\frac{-r_{\mathrm{H}}^2 (D-2) + 2\lambda (D-3)(D-4)}{r_{\mathrm{H}}^2 + 2 \lambda (D-3) (D-4)}\right)\right)\nonumber,\\
&&r_{\mathrm{H}}^2 + (D-3) (D-4) \lambda  =  \left(\frac{r_{\mathrm{H}}^2 + (D-4) (D-5) \lambda}{r_{\mathrm{H}}^2 +2  \lambda (D-3) (D-4)}\right)\left(r_{\mathrm{H}}^2 + 2 (D-2) (D-3) \lambda \right)\nonumber\\
&&+ 2 \lambda   \left(\frac{-r_{\mathrm{H}}^2 (D-2) + 2\lambda (D-3)(D-4)}{r_{\mathrm{H}}^2 + 2 \lambda (D-3) (D-4)}\right)\nonumber,\\
&&\left(r_{\mathrm{H}}^2 + (D-3) (D-4) \lambda \right)\left(r_{\mathrm{H}}^2 + 2\lambda(D-3) (D-4)\right) =  \left(r_{\mathrm{H}}^2 + (D-4) (D-5) \lambda\right)\left(r_{\mathrm{H}}^2 + 2 (D-2) (D-3) \lambda\right)\nonumber\\
&&+ 2 \lambda \left(-r_{\mathrm{H}}^2 (D-2) + 2 (D-3)(D-4) \lambda\right)\nonumber,\\
&& r_{\mathrm{H}}^4 +  r_{\mathrm{H}}^2 \lambda \left(2 (D-3) (D-4) + (D-3) (D-4)\right) + 2 \lambda^2 \left((D-3)^2 (D-4)^2\right)\nonumber\\
&&= r_{\mathrm{H}}^4  + r_{\mathrm{H}}^2 \lambda  \left(2 (D-2) (D-3) + (D-4) (D-5)-2(D-2)\right)\nonumber\\
&&+\lambda^2 \left(2 (D-2) (D-3) (D-4) (D-5)+ 4 (D-3) (D-4) \right),\nonumber\\
&& r_{\mathrm{H}}^4 + 3 r_{\mathrm{H}}^2 \lambda \left(D^2 - 7 D + 12\right) + 2 \lambda^2 \left((D-3)^2 (D-4)^2\right)\nonumber\\
&&= r_{\mathrm{H}}^4  + r_{\mathrm{H}}^2 \lambda  \left(2 D^2 - 10 D + 12   + D^2  - 9 D + 20 - 2D + 4)\right)\nonumber\\
&&+\lambda^2 \left(2 (D-2) (D-3) (D-4) (D-5)+ 4 (D-3) (D-4) \right),\nonumber\\
&& r_{\mathrm{H}}^2 \lambda \left(3 D^2 - 21 D +36)\right)- 3 r_{\mathrm{H}}^2 \lambda \left(D^2 - 7 D + 12\right) \nonumber\\
&&+\lambda^2 (D-3)(D-4)\left(2 (D-2) (D-5)+ 4 \right)-2 \lambda^2 \left((D-3)^2 (D-4)^2\right)=0,\nonumber\\
&&3 r_{\mathrm{H}}^2 \lambda \left(D^2 -7D + 12 -D^2 + 7D -12\right) + 2\lambda^2 (D-3)(D-4) \left(D^2-7D+10+2-(D-3)(D-4)\right)=0,\nonumber\\
&&2\lambda^2(D-3)(D-4)\left((D-3)(D-4)-(D-3)(D-4)\right)=0.
\end{eqnarray}
Hence, as a result, the Smarr relation is satisfied for a generic $D$-dimensional Einstein-Gauss-Bonnet gravity.

\section*{ACKNOWLEDGMENTS}
We thank Kamal Hajian and M.M. Sheikh-Jabbari for useful discussions.


\begin{thebibliography}{10}
\bibitem{Hawking1}
	S.~W.~Hawking,
	Particle creation by black holes,
	Commun. Math. Phys. \textbf{43}, 199 (1975); \textbf{46}, 206(E) (1976).
	
	\bibitem{Visser1}
	T.~Jacobson and R.~M.~Visser,
	Partition function for a volume of space,
	Phys. Rev. Lett. \textbf{130}, 221501 (2023).	
	
	\bibitem{Tavlayan1}
	A.~Tavlayan and B.~Tekin,
	Partition function of a volume of space in a higher curvature theory,
	Phys. Rev. D \textbf{108}, L041902 (2023).	
	
	\bibitem{Bekenstein1}
	J.~D.~Bekenstein,
	Black holes and entropy,
	Phys. Rev. D \textbf{7}, 2333 (1973).
	
	\bibitem{Gibbons1}
	G.~W.~Gibbons and S.~W.~Hawking,
	Action integrals and partition functions in quantum gravity,
	Phys. Rev. D \textbf{15}, 2752 (1977).	
	
	\bibitem{Smarr1}
	L.~Smarr,
	Mass formula for Kerr black holes,
	Phys. Rev. Lett. \textbf{30}, 71 (1973); 30, 521(E) (1973).

	\bibitem{Hajian1}
	K.~Hajian and B.~Tekin,
	Coupling constants as conserved charges in black hole thermodynamics,
	Phys. Rev. Lett. \textbf{132}, 191401 (2024).	
	
	\bibitem{Kastor2}
	D.~Kastor, S.~Ray, and J.~Traschen,
	Smarr formula and an extended first law for Lovelock gravity,
	Classical Quantum Gravity \textbf{27}, 235014 (2010)

	\bibitem{Komar1}
	A.~Komar,
	Covariant conservation laws in general relativity,
	Phys. Rev. \textbf{113}, 034 (1959).	

	\bibitem{Abbott1}
	L.~F.~Abbott and S.~Deser,
	Stability of gravity with a cosmological constant,
	Nucl. Phys. B\textbf{195}, 76 (1982).
  
	\bibitem{Deser1}
	S.~Deser and B.~Tekin,
	Gravitational energy in quadratic-curvature gravities,
	Phys. Rev. Lett. \textbf{89}, 101101 (2002).	

	\bibitem{Deser2}
	S.~Deser and B.~Tekin,
	Energy in generic higher curvature gravity theories,
	Phys. Rev. D \textbf{67}, 084009 (2003).	

 	\bibitem{Hawking2}
    S.~W.~Hawking, C.~J.~Hunter, and M.~Taylor,
    Rotation and the AdS / CFT correspondence,
    Phys. Rev. D \textbf{59}, 064005 (1999).

    \bibitem{Pope1}
    G.~W.~Gibbons, H.~Lu, D.~N.~Page, and C.~N.~Pope,
    Rotating black holes in higher dimensions with a cosmological constant,
    Phys. Rev. Lett. \textbf{93}, 171102 (2004).	

    \bibitem{Pope2}
    G.~W.~Gibbons, H.~Lu, D.~N.~Page, and C.~N.~Pope,
    The general Kerr-de Sitter metrics in all dimensions,
    J. Geom. Phys. \textbf{53}, 49 (2005).

	\bibitem{Myers1}
	R.~C.~Myers and M.~J.~Perry,
	Black holes in higher dimensional space-times,
	Ann. Phys. (N.Y.) \textbf{172}, 304 (1986).	

	\bibitem{Cvetic1}
	M.~Cveti\v{c}, G.~W.~Gibbons, D.~Kubiz\v{n}\'ak, and C.~N.~Pope,
	Black hole enthalpy and an entropy inequality for the thermodynamic volume,
	Phys. Rev. D \textbf{84}, 024037 (2011).	
 
	\bibitem{Hajian2}
	K.~Hajian, H.~\"Oz\c{s}ahin, and B.~Tekin,
	First law of black hole thermodynamics and Smarr formula with a cosmological constant,
	Phys. Rev. D \textbf{104}, 044024 (2021).

    \bibitem{Henneaux1}
    M.~Henneaux and C.~Teitelboim,
    The cosmological constant as a canonical variable,
    Phys. Lett. \textbf{143B}, 415 (1984).

    \bibitem{Chernyavsky1}
    D.~Chernyavsky and K.~Hajian,
    Cosmological constant is a conserved charge,
    Classical Quantum Gravity \textbf{35}, 125012 (2018).

    \bibitem{New1}
    E.~Altas and B.~Tekin,
    Conserved charges in AdS: A new formula,
    Phys. Rev. D \textbf{99}, 044026 (2019).

    \bibitem{New2}
    E.~Altas and B.~Tekin,
    New approach to conserved charges of generic gravity in AdS spacetimes,
    Phys. Rev. D \textbf{99}, 044016 (2019).

 	\bibitem{Deser3}
	S.~Deser, I.~Kanik, and B.~Tekin,
	Conserved charges of higher D Kerr-AdS spacetimes,
	Classical Quantum Gravity \textbf{22}, 3383 (2005).

	\bibitem{Brown2}
	J.~D.~Brown and C.~Teitelboim,
	Dynamical neutralization of the cosmological constant,
	Phys. Lett. B \textbf{195}, 177 (1987).

    \bibitem{Caldarelli1}
    M.~M.~Caldarelli, G.~Cognola, and D.~Klemm,
    Thermodynamics of Kerr-Newman-AdS black holes and conformal field theories,
    Classical Quantum Gravity \textbf{17}, 399 (2000).

	 \bibitem{Kastor1}
    D.~Kastor, S.~Ray, and J.~Traschen,
    Enthalpy and the mechanics of AdS black holes,
    Classical Quantum Gravity \textbf{26}, 195011 (2009).

    \bibitem{Kubiznak1}
    D.~Kubiznak and R.~B.~Mann,
    P-V criticality of charged AdS black holes,
    J. High Energy Phys. \textbf{07} (2012) 033.


	\bibitem{Kubiznak2}
    D.~Kubiznak, R.~B.~Mann, and M.~Teo,
    Black hole chemistry: Thermodynamics with Lambda,
    Classical Quantum Gravity \textbf{34}, no.6, 063001 (2017).

    \bibitem{Dolan1}
    B.~P.~Dolan,
    \textit{Where Is the PdV in the First Law of Black Hole Thermodynamics?} (2012),
    https://doi.org/10.5772/52455.

	\bibitem{Boulwere1}
	D.~G.~Boulwere and S.~Deser,
	String-generated gravity models,
	Phys. Rev. Lett. \textbf{55}, 2656 (1985).	
 
	\bibitem{Tahsin1}
	M.~G\"urses, T.~\c{C}.~\c{S}i\c{s}man, and B.~Tekin,
	Is there a novel Einstein\textendash{}Gauss\textendash{}Bonnet theory in four dimensions?,
	Eur. Phys. J. C \textbf{80}, 647 (2020).
	
	\bibitem{Tahsin2}
	M.~G\"urses, T.~\c{C}.~\c{S}i\c{s}man, and B.~Tekin,
	Comment on Einstein-Gauss-Bonnet gravity in 4-dimensional space-time,
	Phys. Rev. Lett. \textbf{125}, 149001 (2020).
 
	\bibitem{Mann1}
	S.~Haroon, R.~A.~Hennigar, R.~B.~Mann, and F.~Simovic,
	Thermodynamics of Gauss-Bonnet-de Sitter black holes,
	Phys. Rev. D \textbf{101}, 084051 (2020).
 
	\bibitem{Liberati1}
	S.~Liberati and C.~Pacilio,
	Smarr formula for Lovelock black holes: a Lagrangian approach,
	Phys. Rev. D \textbf{93}, 084044 (2016).

	

	
\end{thebibliography}
\end{document}